\documentclass[preprint]{elsarticle}   
\usepackage{graphicx}  
\usepackage{dcolumn}   
\usepackage{bm}        
\usepackage{amssymb}   
\usepackage{verbatim}  
\usepackage{epstopdf}
\usepackage{color}
\usepackage{soul}
\usepackage{amsmath}
\usepackage{epstopdf}
\usepackage{bbold}
\usepackage{url}

\newcommand{\ket}[1]{\ensuremath{\left| #1 \right\rangle}}
\newcommand{\bra}[1]{\ensuremath{\left\langle #1 \right|}}

\newcommand{\ave}[1]{{\langle{#1}\rangle}}
\newcommand{\tr}{{\rm Tr}}
\newcommand{\dyad}[2]{{\ket{#1}\!\!\bra{#2}}}

\newcommand{\beq}{\begin{equation}}
\newcommand{\eeq}{\end{equation}}
\newcommand{\bea}{\begin{eqnarray}}
\newcommand{\eea}{\end{eqnarray}}

\newcommand{\eq}[1]{{(\ref{#1})}}
\newcommand{\commentout}[1]{{}}

\newcommand{\half}{{\hbox{$\frac{1}{2}$}}}

\newcommand{\bE}{{\bf E}}
\newcommand{\br}{{\bf r}}

\newcommand{\bd}{{\bf d}}

\newcommand{\code}[1]{{\tt  #1}}

\newcommand{\bmu}{\boldmath\hbox{$\mu$}}
\begin{document}

\title{The Software Atom}

\author{Juha Javanainen${}^*$}
\address{Department of Physics, University of Connecticut, Storrs, Connecticut 06269-3046, USA\\
\flushleft{${}^*$\rm email: jj@phys.uconn.edu}}
\begin{abstract}
By putting together an abstract view on quantum mechanics and a quantum-optics picture of the interactions of an atom with light, we develop a corresponding set of C++ classes that set up the numerical analysis of an atom with an arbitrary set of angular-momentum degenerate energy levels, arbitrary light fields, and an applied magnetic field. As an example, we develop and implement perturbation theory to compute the polarizability of an atom in an experimentally relevant situation.
\end{abstract}
\maketitle

\section{Introduction}
\begin{figure}[b]
\vspace{-12pt}
\center\includegraphics[width=0.65\columnwidth]{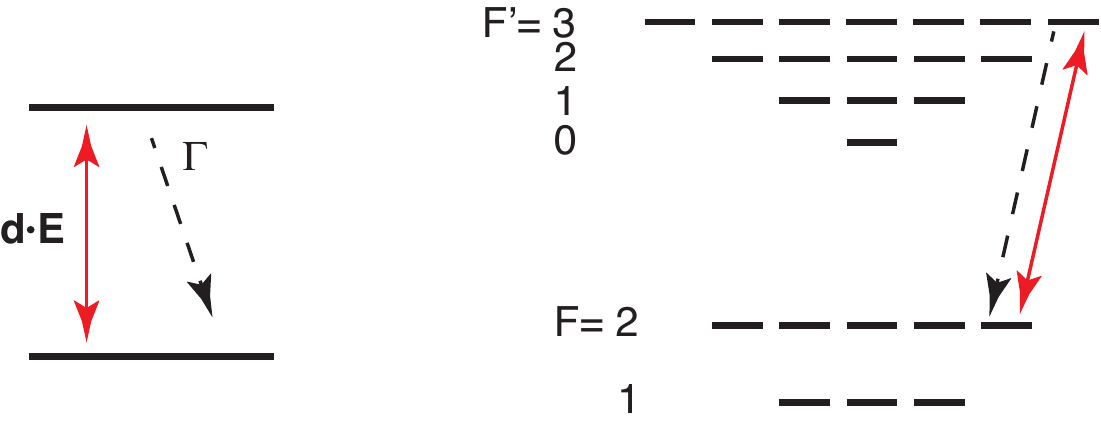}
\vspace{-12pt}
\caption{Left: The two-level system. Also shown symbolically are the dipole-coupled transitions driven by
the external light, and spontaneous emission taking the atom down from the upper to the lower level. Right: Schematic representation of the hyperfine structure of the $D_2$ line in ${}^{87}$Rb. The light-driven and spontaneous transitions between the ``stretched'' states $F=2,\,m_F=2$ and $F'=3,\,m_F'=3$ are also illustrated.}
\label{LEVEL_SCHEMES}
\end{figure}

In the eponymous two-level atom the entire atomic level structure is reduced to two states with a dipole coupling between them. An external light couples to the dipole moment, and drives transitions between the states or ``levels.'' Transitions between the two states and possibly also to states outside of the two-state system associated with spontaneous emission of photons may also be included; see Fig.~\ref{LEVEL_SCHEMES}, left. These concepts have been thoroughly discussed in many quantum-optics textbooks~\cite{WAL94,MEY99}, and have proven enormously valuable in the analysis of atom-field interactions.  For instance, the two-level system is the reigning paradigm of the physical implementations of quantum information systems.

However, Nature does not necessarily cooperate. Atomic energy levels have angular momentum degeneracy except if the total angular momentum is zero, but there is no dipole transition between two states with zero angular momenta.  Consult Fig.~\ref{LEVEL_SCHEMES}, right, for a highly stylized representation of the hyperfine structure of the $D_2$ line of ${}^{87}$Rb, an atomic species and transition commonly employed in modern atomic physics experiments. The total number of quantum states, including the Zeeman states in the hyperfine levels, is 24. Now, if the driving transition has the proper $\sigma^+$ circular polarization, there is a closed transition between the stretched states, ground- and excited-level hyperfine states $F=2,\,m_F=2$ and $F'=3,\,m_F'=3$, in such a way that the selection rules for dipole transitions prevent the atom from leaving this two-state system; and there is a pathway from every other state to the stretched states with dipole-allowed light-driven and spontaneous transitions. The atom thus invariably (in principle)  gets optically pumped to the two stretched states, and a two-level system is realized. But there are complications, and the experiments often add a second ``repumper'' laser to help things along.

Numerous are the cases when experimental results from laboratories are qualitatively discussed in terms of two-level models, but more enlightenment is sought from computations using multistate models. Besides, there are also experiments in which level systems with more that two states are a deliberate objective. For instance, the multitude of quantum states may be an advantage in laser cooling, leading to much lower atomic temperatures than would be possible for a two-level atom.

Well over 20 years ago we wrote a C code that in principle could manage an arbitrary atomic level structure under an arbitrary light field, and used it to study problems such as laser cooling in situations where angular momentum degeneracy is essential~\cite{JAV91,JAV92,JAV92_EPL,JAV94,WALLACE94,JAV94_JPB}. We  also have more modest C programs for use in M\"ossbauer spectroscopy~\cite{TIT96}, with the main enhancement over the atomic-physics scheme being that we allow arbitrary multipole transitions in addition to dipole transitions. Since then C++~\cite{STR13} has all but superseded C. Moreover, with the interest in quantum information, quantum measurements and feedback~\cite{NIE00,WIS10}, deep understanding of the structural aspects of quantum mechanics is more common these days than it was back then. That is why we undertook a complete rewriting of the C code in C++.

The idea of the new Software Atom is to abstract the relevant concepts of quantum mechanics and atomic physics, and map them to software objects. We describe here the relevant mathematics and physics background, discuss the corresponding programming principles and software, and illustrate with code snippets. Finally we discuss briefly an application to physics related to experimental and theoretical analyses of light scattering from a small, dense cloud of low-temperature atoms.

Our emphasis is really on philosophy, namely, the synergy between physics, mathematics, and coding. We do not offer any comprehensive and optimized end-user programs. However, an operational version of the C++ code with a main program and thorough internal commenting is available at~\cite{JAV16}. 

\section{Vectors, operators, and so on}
\subsection{Primitives of quantum mechanics}
State vector $\ket v$ is the elementary concept of state in quantum mechanics. An array of complex numbers $v_i$, numerically, a complex vector, is a natural representation of a state vector. Along with the vectors comes the inner product of two vectors that can also be implemented numerically in a natural way.

To deal with vectors, we have coded a template vector-matrix class. Vectors and matrices come with the usual algebra implemented by overloading C++ operators. For instance, if \code{a} and \code{b} are complex numbers, \code{u}, \code{v}, and \code{w} are complex vectors, and $\code{A}$ is a complex matrix, then \code{w=a*u+b*w} and \code{v=A*u} do what you would expect them to do, \code{inner(u,v)} takes the inner product of two vectors, and \code{u.normalize()} normalizes the vector~\code{u} to unity. Even some implicit type conversions are implemented, so if \code r  is a real vector and \code c is a complex vector, the operation \code{c=r} is legal. For the present purpose we have defined, using \code{typedef}, both the scalar type \code{CMPLX} as \code{complex<double>},  and the shorthands \code{RealV}, \code{ComplexV}, \code{RealM} and \code{ComplexM} for real and complex vectors and matrices..

Our matrices and vectors known about their dimensions, which are checked on every attempt to do algebra and an exception is thrown if the dimensions are not compatible. This may not look efficient, but it helps enormously to catch programming errors. On the scale of computational problems the Software Atom is light work, so the programmer's time is much more costly than the CPU time. That is why the checking of the dimensions cannot be bypassed in the present version.

On the next level of quantum mechanics are the (linear) operators. Their primary implementation is in terms of the matrices of the same vector-matrix class that we have already mentioned. Our Software Atom also uses internally another implementation of operators in which the matrix elements of the operator are stored as a linked list maintained by using the C++ Standard Template Library. One element in the list consists of the indices of the matrix and the corresponding matrix element.

A particularly prominent operator in quantum mechanics is the density operator $\rho$, a positive operator with unit trace. It is a generalization of the usual state-vector state of a quantum system. In fact, if there exists a normalized state vector $\ket\psi$  such that $\rho=\dyad{\psi}\psi$, the density operator is said to be a pure state. The expectation value of the quantum mechanical observable is given by $\ave A = \tr(\rho A)$.

We build on top of  the vector-matrix package, bringing in additional layers of quantum mechanics and  relevant atomic physics. First, we allow a vector implementation of operators, $A_k$ with just one index, with a one-to-one mapping between the indices $k$ and the pairs of vector indices $ij$. Although this choice is hidden from the user, in the present code the matrix-form operator $A_{ij}$ is in fact stored as the exact same array of complex numbers  as the corresponding vector-form operator $A_k$, in the C style row-major order.  We also offer constructs such as outer product, construction of operators from vectors as in the Dirac notation dyad $\dyad{\psi}{\phi}$.

The vector-matrix dichotomy together with our strict checking of dimensions of vectors and matrices necessitates explicit conversions between the two operator types. The function \code{convert} does it both ways, depending on the type of the argument. In quantum mechanics with density operators trace also plays a prominent role, so we have two functions \code{Trace} to deal with it for operators represented as vectors.

The next level of abstraction in quantum mechanics is operators acting on operators. For instance, think about the Liouville-von Neumann equation of motion of the density operator in a closed system with the Hamiltonian $H$,
\beq
\frac{d}{dt}\rho = -\frac{i}{\hbar}[H,\rho]\,.
\label{SOPEX}
\eeq
The right-hand side clearly presents a linear operator, a superoperator, that takes the operator $\rho$ as an argument and turns out another operator $(-i/\hbar)[H,\rho]$ as the result.
At the basic level the superoperators are collections of complex numbers ${\cal L}_{ij,kl}$ labeled by four vector indices, and the action of the superoperator ${\cal L}$ on the operator $A$ produces the operator ${\cal L}A$ whose matrix elements are $({\cal L}A)_{ij}=\sum_{kl} {\cal L}_{ij,kl}A_{kl}$.

If we think of operators as vectors, then superoperators can be thought of as matrices and represented using the existing vector-matrix class. In this way the linear algebra of superoperators is automatically implemented. However, as it turns out that in this problem area matrices holding superoperators are generally sparse, we also have another implementation as a linked list based on the C++ Standard Template Library. We have not coded the linear algebra for this representation of the superoperators, but the linked-list superoperators know how to act on operators represented both as matrices and as vectors.

\subsection{Quantum mechanics wrapped in an atom}

In the end, though, a typical user of the Software Atom needs to see relatively little of the underlying vector-matrix algebra we have described until now. This is because it is wrapped inside abstractions mimicking atomic physics. At the core there is the class \code{general\_atom}, which encapsulates the description of an atom.

In the vein of quantum optics we think of an atom as an arbitrary collection of angular-momentum degenerate levels of Zeeman states with dipole couplings between the levels as appropriate. Correspondingly, we have a class \code{level} for each level. One specifies a level by giving the degeneracy equal to $2J+1$, with $J$ being the angular momentum of the level, and the Lande $g$ factor that is needed if one is to allow a magnetic field to act on an atom; if not specified, the default value is 0. However, we use C++ access rules to disallow direct manipulation of levels in an atom. Instead, to construct an atom, a user would start by taking an empty atom and adding levels to it, as here a level with $J=1$:
\code{\vspace{5pt}\noindent\\
general\_atom atom;\\
size\_t deg\_e = 3;\\
double Lande\_e = 0.653;\\
atom.add\_level("e", deg\_e, Lande\_e);\\
\vspace{-5pt}}\\
The string literal \code{"e"} will internally turn into a \code{string} that is subsequently used to refer to this level.

From the quantum optics point of view, though, the atom generated so far is uninteresting in that it does not couple to the external light. The next step is to add the appropriate dipole couplings between the levels. Suppose we have two levels \code{g} (``ground'') and \code{e} (``excited'') defined for an atom, then we add a dipole coupling between the levels as here,
\code{\vspace{5pt}\noindent\\
double D = 1.0;\\
double gamma = 1.0;\\
atom.add\_coupling("g",  "e",  D,  gamma);\\
\vspace{-5pt}}\\
\code{D} is the reduced dipole matrix element defined so that the dipole matrix element between magnetic substates equals $D$ times the corresponding Clebsch-Gordan coefficient, and \code{gamma} is the half width half maximum linewidth for the transition between these levels in the absence of any other levels. Note that the order of \code{"g"} and \code{"e"} matters: \code{e} is taken to be higher in energy so that spontaneous decay proceeds in the direction \code{e}$\rightarrow$\code{g}. In a maybe counterintuitive design decision explained below, the levels themselves do not know what their energies are.

The atom thus far constructed is an organizational template. It holds a description of the levels and the transitions between them, and some static data such as the relevant Clebsch-Gordan coefficients. An atom also stores the data needed for a quantitative description of spontaneous-emission transitions that might occur in it. Finally, an atom presents a facility to make any state vector using a symbolic representation of levels and Zeeman states without having to know explicitly the vector representation of the state. Thus, given the complex vector type \code{ComplexV}, we could make a state vector that has the atom in a particular level $g$ and Zeeman state $m=2$,
\code{\vspace{5pt}\noindent\\
ComplexV state(atom.get\_size());\\
double m = 2;\\
double value = 1.0;\\
atom.add\_to\_state("g",  m,  value,  state);\\
\vspace{-5pt}
}\\
The index for the Zeeman state $m=2$ is a \code{double} to accommodate half-integers; if the given \code{m} is not close to an integer or a half-integer or is otherwise invalid, an error message and an abort result.

Next we have the atomic operator class  \code{atomic\_operator}. The constructor  \code{atomic\_operator(general\_atom\& atom)} stores a reference to \code{atom}, which is then used to construct an operator acting on the atom. Objects of the class \code{general\_atom} know how to calculate their expectation values in both state vector states and density operator states, and also how to construct certain superoperators corresponding to the operator, such as the one the right-hand side of \eq{SOPEX}. 

In fact, an object of the type \code{atomic\_operator} would, however, be a zero operator. The substance is in the derived classes. We have developed a surprisingly small collection that has so far answered all of our questions, such as magnetic moment operator and unit operator for each level, and the raising and lowering parts of the dipole operator between the levels. For instance, given two levels $g$ and $e$, $g$ the one being the one lower in energy, the raising operator is
\beq
{\bd}^+ = D\sum_{m_g,m_e,\sigma} \hat{\bf e}^*_\sigma C^{J_e,m_e}_{J_g,m_g;1,\sigma} \dyad{e,m_e}{g,m_g},
\label{DIPOP}
\eeq
and the lowering operator $\bd^-$ is its hermitian conjugate. Here $\hat{\bf e}_\sigma$, $\sigma=-1,0,1$ are the usual circular unit operators, and $C$ stand for the appropriate Clebsch-Gordan coefficients. To be sure, we also have a catch-all class that can be constructed to contain any operator with arbitrary matrix elements specified in terms of the symbolic level designators and magnetic quantum numbers.

On the programming side, we have added safeguards against potential problem situations. For example, once any object that depends on the mapping of atomic states to complex vectors and matrices is instantiated, the atom gets locked and can no longer be modified. There are also member functions in the class \code{general\_atom} to print out detailed descriptions of vectors, operators and superoperators using the symbolic level-Zeeman state description to label the states. These have proven invaluable for debugging. Finally, access control is used extensively to prevent potentially dangerous direct manipulations of the innards of the objects.

To sum up what we have so far, we have at our disposal a machine to construct an arbitrary dipole coupled atomic level structure and a ready-made collection of the most useful operators corresponding to the observables of the atom. Besides, we have facilities to construct any state vector and any operator symbolically without having to know how they are represented as complex vectors and matrices. 

\subsection{Atom and external fields}
Nonetheless, the atom that we have so far is still of little use in quantum optics style problems: Neither the light fields driving the atom nor a possible applied magnetic field are accounted for. This is the task for the class \code{atom\_and\_fields}. It is derived from \code{general\_atom}, and the member functions \code{add\_level} and \code{add\_coupling} also work for  \code{atom\_and\_fields} objects. In fact, one does not need an explicit \code{general\_atom} object at all.

When an atomic transition with the frequency $\omega$ is driven by an electric field with a (dominant) frequency $\bar\omega$, the ordinary practice in quantum optics is to make a unitary transformation of the quantum mechanics to what is known as a rotating frame, and perform the rotating-wave approximation~\cite{WAL94,MEY99}. Three things effectively happen. First, a physical electric field of the form $\bE(\br,t) = \half[\bE^+(\br,t)e^{-i\bar\omega t}+{\rm c.c.}]$ gets transformed to a slowly varying (or possibly stationary) field $\half \bE^+(\br,t)$.  Second, the characteristic frequency of the atomic level is set at $\Delta$ where the detuning is
\beq
\Delta=\omega-\bar\omega.
\eeq
 Be warned, in quantum optics it is more common to define the detuning as $\bar\omega-\omega$. Third, there will be dipole couplings of the electric field to the atomic dipoles. We split the dipole operators in two parts, positive frequency part $\bd^+$, see~\eq{DIPOP}, which acts to transfer atoms from the states that are actually (in real terms) lower in energy to states with higher energy, and its Hermitian conjugate $\bd^-$.  The end result is a dipole coupling term in the Hamiltonian of the form
$
-\half \bE^+(\br,t)\cdot\bd^+ + {\rm h.c.}
$
In the code we write all vectors in terms of the usual cartesian components, numbered 0, 1, and 2 for $x$, $y$ and $z$. Given the complex-valued three-dimensional vectors, we occasionally need an inner product that is appropriate for complex vectors. This circumstance has spurred various notational conventions in the literature. Here we use a dot product that means the same as the usual dot product of three-dimensional {\em real\/} vectors, and, when needed, write the conjugates that come with the complex inner product explicitly.

Let us now assume that an \code{atom\_and\_fields} object \code{af} has already been defined and furnished with two levels, ground level \code{g} and excited level \code{e}. To gain access to the detuning, i.e., apparent energy of the excited level, we say
\code{\vspace{5pt}\noindent\\
det\_id id = af.variable\_detuning("e");\\
\vspace{-5pt}}\\
The purpose of this line is threefold. First, it declares that the energy of the excited state \code{e} may be varied. Second, it gives an object~\code{id} of the type \code{det\_id} that acts as the handle to the detuning to be varied. Third, it adds to the Hamiltonian of the atom-field system a term corresponding to an apparent transition frequency $\Delta$ of the form
\beq
\frac{H_{e}}{\hbar} = \Delta \mathbb{1}_e,
\eeq
where $\mathbb{1}_e$ is the unit operator in the manifold of the states belonging to the level $e$. We would subsequently change the apparent energy of the excited state to the value \code{delta} by saying
\code{\vspace{5pt}\noindent\\
af.set\_det(id, delta);\\
\vspace{-5pt}}\\
The default is $\Delta=0$ and only energy differences matter anyway, so in a two-level example we may leave the ground state alone.

It may be that the system presents couplings between several levels, and in a complicated configuration. For instance, we may have a transition from a ground state $g$ to an excited state $e$ and then further up to a state $E$; or down from $e$ to another level $G$. There are standard procedures in quantum optics how to set the detunings in each such case. For instance, you could fix the detuning of one of the levels at zero and calculate how much multiphoton transitions starting from this level and with the given light frequencies would fall below resonance with each of the other levels. This value equals the detuning; if the multiphoton transitions put you above the resonance, the detuning is simply negative. To put this formally, let us study sequential transitions from a reference level with the characteristic frequency $\omega_0$ to the levels $i$ with the characteristic frequencies $\omega_i$ with lasers whose frequency is denoted by $\bar\omega_1$ for the transition from the level 0 to 1, and so on. Then the detuning for the level $i$ would be
\beq
\Delta_i = (\omega_i - \omega_0) - \sum_{j=1}^i {\rm sgn}(\omega_i-\omega_{i-1})\bar\omega_i\,,
\eeq
where ${\rm sgn}$ is the signum function.
There is also an added condition of consistency: If there are two (or more) paths via light-induced transitions between two levels, the detuning should come out the same along both paths. Our earlier general-atom scheme~\cite{JAV91} automatically checked that this rule is satisfied, but here the check is not implemented.  If the rule is not satisfied, it is not possible to find a rotating frame that would eliminate the oscillation frequencies of all driving fields completely from the description. The Software Atom could be used to deal even with this situation, but since we have never run into it and it seems extremely unlikely in practice anyway, we will not describe how.
 
Management of a light field coupling to a pair of levels works analogously. Suppose we want to set the electric field with the positive frequency part $\bE^+$ to act on the transition between the same ground and excited levels \code{g} and \code{e} as in the preceding examples. To do it, we would first prepare a three-dimensional complex vector \code{E\_plus} with the $x$, $y$, and $z$ components of $\bE^+$ then program as follows,
\code{\vspace{5pt}\noindent\\
E\_id iE = af.variable\_E("g", "e");\\
af.set\_E(iE, E\_plus);\\
\vspace{-5pt}}\\
This would add a dipole coupling between the levels in the Hamiltonian,
\beq
H_{ge} = -\half (\bd^+\cdot\bE^++\bd^-\cdot\bE^-).
\eeq
The object \code{iE} of the type \code{E\_id} is a handle that keeps track of the transition for which the field is changed to the value \code{E\_plus}. The negative-frequency component of the electric field is automatically deduced from the positive-frequency component, and is used where needed.

There may also be a global magnetic field $\bf B$ acting on the atom that couples to the magnetic moment operators of the atom  $\bmu$, sum of the magnetic moment operators of all levels, and adds the term to the Hamiltonian
\beq
\hbox{$H_B$}=-\bmu\cdot\bf B\,\hbox{.}
\eeq
There is no need to specify what level or transition the magnetic field acts on and correspondingly no handle class, but for consistency we still require that the magnetic field be declared as variable. To set the magnetic field to the value $\bf B$, we first prepare a three-dimensional  real vector \code{B} so that the $z$, $y$, and $z$ components of the magnetic field are in \code{B[0]}, \code{B[1]}, and \code{B[2]}, and write
\code{\vspace{5pt}\noindent\\
af.variable\_B();\\
af.set\_B(B);
\vspace{-5pt}}\\

At this point we have constructed the Hamiltonian governing the atom under the given fields and detunings. In fact, \code{af.get\_H} would return a constant reference to the Hamiltonian that one could use to integrate the time evolution of the state vector of the atom, and likewise for the superoperator corresponding to the Hamiltonian evolution of the density operator. These operators are set up by the declarations that enable detunings and electric and magnetic fields to be varied, and are adjusted with a minimal number of operations every time the field values or the detuning are changed. For instance, if the detuning of a level is changed, the elements in the Hamiltonian that do not depend on this detuning are not recomputed or otherwise touched. 

However, this is not yet the end of the line. In real atoms there are also spontaneous-emission processes in which the atoms make transitions from upper to lower levels as a result of the dipole coupling to the vacuum of the electromagnetic field. Another way of viewing the same is that the electric field originating from a transition couples back to the transition that sent it and causes the atom to make a spontaneous transition. It is also possible that the field from one transition couples to another, and peculiar cross-couplings between spontaneous-emission processes emerge. These are a fact of life in transitions between the Zeeman states of two angular-momentum degenerate levels, since light emitted in one transition is close to resonance with the other transitions and can couple strongly to them. We have a prescription to take into account all of these spontaneous-emission terms~\cite{JAV91}, and we do so here. It is possible to have spontaneous cross-couplings in transitions between different levels, too, especially if the transition frequencies are close~\cite{JAV92_SGC}. However, the symmetries in atomic physics generally eliminate them, and they manifest only by unlikely accident. We had the possibility of spontaneous couplings between different levels included in our C version of the software atom, but no more here.

The bottom line is that there exists a superoperator $\cal G$ that describes the effect of spontaneous emissions. Taking this into account, the evolution of the density operator is given in terms of yet another superoperator $\cal L$ as
\beq
\frac{d}{dt}\rho = {\cal L}\rho = \frac{1}{i\hbar}[H,\rho] + {\cal G}\rho\,.
\label{EQM}
\eeq
An object of class \code{atom\_and\_fields} forms all of the relevant superoperators when needed.
One typical consequence of the ensuing time evolution is that even if the state were pure at some particular time, in general it does not remain so at all times and has to be described by a density operator instead of a state vector.

We now have the objects available that generate the time evolution of the density operator as per the superoperators on the right-hand side of~\eq{EQM}. We could wrap the right-hand side inside a differential-equation solver, and find the evolution of the density operator and whatever quantities are deemed interesting for the prevailing purpose even for explicitly time dependent external fields. However, at present we have not implemented any time dependent calculations in the C++ version of the Software Atom. Instead, we have developed a small demonstration project about the steady state: We calculate the polarizability tensor of an atom. Before proceeding to our example, we note that others have developed codes for essentially the same purpose~\cite{ZEN15}. In comparison, our method to obtain the polarizability is inherently more general, e.g., we have no restrictions on the direction of the magnetic field. On the other hand, we do not offer a front end to set up the level structure for many commonly used isotopes of alkali atoms, nor a back-end to, say, fit the calculation results to experiments.
 
\section{Example: Polarizability of ${}^{87}$Rb}

In Refs.~\cite{PEL14,Jenkins_thermshift,JEN16} the authors compare basically exact simulations of light scattering with experimental results obtained in very small atomic samples, and find in both ways that the shift of the fluorescence line shape (fluorescence intensity as a function of laser tuning) with the density of the sample is much smaller than one expects from the traditional local-field correction arguments of electrodynamics~\cite{JAV14}. For the present purposes the point is that the ground state of the atom used, ${}^{87}$Rb, has the total angular momentum $F=2$, and is thus five-fold degenerate. The excited level of the transitions has $F'=3$, and seven-fold degeneracy. A magnetic field was also applied in the experiments.  We have coded the same ${}^{87}$Rb simulations as the group of J. Ruostekoski~\cite{PEL14} independently (without ever seeing their code) using the Software Atom, and in collaboration we have verified that, to within statistical error, the results are the same. As an aside the question arose, what is the polarizability of the ${}^{87}$Rb atom in an arbitrary external magnetic field, and for an arbitrary but known steady state that the atom may have in the presence of the magnetic field.

The evolution of an atom with spontaneous emissions is damped and, for constant fields and detunings usually (though not necessarily always) leads to a steady state. We start from the equation of motion of the density operator
\beq
\dot\rho = {\cal L}\rho,
\label{LEQM}
\eeq
where the Liouvillean superoperator $\cal L$ includes both the external fields that we wish to take into account while determining the steady state, and all of spontaneous emissions. The steady state satisfies $\dot\rho={\cal L}\rho=0$, but precisely because there is a steady state the matrix $\cal L$ has a zero eigenvalue and the solution to the set of linear equations ${\cal L}\rho=0$ cannot be unique. The standard fix is to replace one of the equations in the set ${\cal L}\rho=0$ with a condition specifying that $\tr\rho=1$. This works as long as the steady state is unique, but it is not always so. Unfortunately it also appears that, if the steady state is not unique, in general it is not possible to find out which steady state gets realized without integrating Eq.~\eq{LEQM} from the initial state all the way to the final steady state. 

Our approach to the a priori unknown number of steady-state solutions is to resort to singular value decomposition (SVD)~\cite{NUMRES}. As is, of course, very well known, in the matrix representation the superoperator $\cal L$ may be decomposed in the form
\beq
{\cal L} = U D V^\dagger,
\eeq
where $U$ and $V$ are unitary matrices, and $D$ is a diagonal matrix with nonnegative elements.  Numerically, we embed SVD from CLAPACK~\cite{CLAPACK} into the Software Atom.

The steady states are distinguished by ``very small'' diagonal elements $D_{ii}$, $i\in\cal I$, of the matrix $D$. Numerically, we replace the elements deemed ``very small'' with zeros. The subspace ${\cal N}$ spanned by the corresponding columns of $V$, $v_{i}$, $i\in\cal I$, is the null space of the matrix $\cal L$, and all steady states live in ${\cal N}$. Steady states must be represented by positive density operators with unit trace, and not all members of the subspace $\cal N$ qualify; but being in $\cal N$ is a necessary and sufficient condition for a valid density operator to be a stationary state.

Given a properly initialized and parameterized \code{atom\_and\_fields} object, the manifold of the steady states would be found by programming
\code{\vspace{5pt}\noindent\\
vector<ComplexV> ss;\\
af.find\_steady\_state(ss);\\
\vspace{-5pt}}\\
If the length of the vector \code{ss} (as returned by \code{ss.size()}) is one, the program has decided that the steady state is unique.  Suppose we then have, say, an object \code{mu\_operator\_z} from the class \code{magmom} containing the $z$ component of the magnetic moment, the expectation value of the magnetic moment in such a steady state would be stored to a newly declared variable \code{mu\_z} by writing
\code{\vspace{5pt}\noindent\\
double mu\_z = real(mu\_operator\_z.expt\_val(ss[0]));\\
\vspace{-5pt}}\\
It is \code{real} since the expectation value of even a hermitian operator usually comes out with a small imaginary part originating from numerical errors.

Continuing in this same vein, the column vectors $v_{j}$, $j\notin\cal I$, span the orthogonal complement  ${\cal N}_\perp$ of the null space. The corresponding column vectors of the matrix $U$, $u_j$, span the range $\cal R$ of the superoperator $\cal L$, all possible operators that can be reached by acting on an arbitrary operator with $\cal L$. Now, ${\cal L}: {\cal N}_\perp\rightarrow {\cal R}$ is a one-to-one mapping of the orthogonal complement of the null space to the range. The inverse of this mapping $\bar{\cal L}:{\cal R}\rightarrow {\cal N}_\perp$, which we call the pseudoinverse of $\cal L$, is obtained as $\bar{\cal L}=V \bar D U^\dagger$, where $\bar D$ is the diagonal matrix with the diagonal elements equal to zero if the corresponding element $D_{ii}$ is zero, for $i\in\cal I$, and otherwise $\bar D_{jj}=1/D_{jj}$. Since the matrix $\cal L$ has a zero eigenvalue, it cannot have a true inverse; but if $v\ne0$ is in the range $\cal R$ of $\cal L$, then $u=\bar{\cal L}v$ is the unique vector outside the null space  $\cal N$ with the property that $v={\cal L}u$

The goal of our mathematical analysis as well as the corresponding code is to device a framework that can find the state of an atom under certain physical/mathematical/computational conditions perturbatively using singular-value decomposition. To be explicit, we phrase the discussion in terms of a ${}^{87}$Rb atom under the experimental conditions of Refs.~\cite{PEL14,Jenkins_thermshift,JEN16}, but the perturbative methods in itself is very general.

We assume that initially there is no light present, so that the steady-state density operator lives  in the manifold spanned by the $F=2$ ground states of the degenerate two-level system. In our example we  specify the steady state manually as it would presumably come out of the preparation of the atoms in the experiment, namely with equal probability in the three lowest-energy states in the given applied magnetic field. Denote the steady state by $\rho_0$. We then add a small electric field $\lambda\bE^+$. Here $\lambda$ is a formal parameter to count orders in perturbation theory. The calculations are carried out in the asymptotic limit $\lambda\rightarrow0$, but the final results are expressed by setting $\lambda=1$. We  then have an added perturbation Hamiltonian denoted by $\lambda H'$, and a perturbation superoperator $\lambda\,\cal K$. Together the perturbation and the existing spontaneous-emission damping bring about a change in the state by $\lambda\rho_1$, so the atom develops an induced dipole moment that turns out to be $\bd = \tr(\rho_1 {\bf d}^-)$. With caveats that we will discuss in detail as we go along, for a weak electric field the dipole moment is linear in the field, ${d_i} = \sum_j \alpha_{ij} E^+_i$. The question is, what is the polarizability tensor $\alpha_{ij}$?

Corresponding to various physical processes that may happen in the atom, the mathematics can get quite challenging. For instance, no matter how small the amplitude of the probing light is, given enough time it may optically pump the atom to a unique steady state that depends on the polarization of the probe but, in the limit of a small amplitude of the probing light not at all on the amplitude. In perturbation theory this would be a correction of the order $\lambda^0$, and would not give an induced dipole moment linear in the driving light. To study the weak-field response, we both retain the interaction strength parameter $\lambda$, and add a mathematical process in which we turn on the probing field on exponentially as $e^{\eta t}$, $\eta>0$, over $t\in(-\infty,0]$. This means that the perturbation is assumedly turned on adiabatically. We formally compute $\rho(0)$ given the initial condition $\rho(-\infty) = \rho_0$, then let $\lambda\rightarrow0$, and finally $\eta\rightarrow0$. In other words, we first take the limit of very low light intensity, then allow a formally infinite time for the optical response to get established.

We thus have an evolution equation for the density operator
\beq
\dot\rho = ({\cal L} + \lambda e^{\eta t} {\cal K})\rho,
\label{PERTEV}
\eeq
and write the solution as a perturbation series, 
\beq
\rho = \rho_0 + \lambda\rho_1+\ldots\,.
\label{PERTSER}
\eeq
The perturbation theory is substantially more involved than, say, in ordinary quantum mechanics, and is detailed in the Appendix. The summary for the present purposes is the following. Suppose the following assumptions hold true:
\begin{itemize}
\item[A1.] We have at our disposal a steady-state density operator $\rho_0$ that satisfies $ {\cal L}\rho_0=0$.
\item[A2.] $ {\cal K}\rho_0\in{\cal R}$.
\item[A3.] We consider the perturbation-induced expectation value of an operator $A$ whose expectation value in any non-perturbed steady state vanishes. More specifically, $\tr (\rho A)=0$, $\rho\in{\cal N}$.
\commentout{\item[3.] ${\cal L}{\cal K}\rho_0\in{\cal N}_\perp$.  This condition tells us that a slightly perturbed state has no stationary component, i.e., the perturbation puts all of the steady state back in play.}
\end{itemize}
Then the system develops a perturbation to the steady state that is linear in the amplitude of the driving light, and the corresponding perturbation-induced expectation value
\beq
\rho_1 = -\bar{\cal L}{\cal K}\rho_0,\quad \ave{A} = \tr(\rho_1A),
\label{FINALRESULT}
\eeq
where  $\bar{\cal L}$ is the pseudoinverse of the Liouvillean. 

To provide a tangible example of coding, we have wrapped the perturbative process into a class \code{pert\_th\_1}. This class even comes with tests of whether the assumptions A1-A3 are valid, i.e., if the perturbation theory is valid as a matter of principle. In the small main program that we provide with the C++ classes~\cite{JAV16} we present a computation of the susceptibility tensor for the ${}^{87}$Rb atom of Refs.~\cite{PEL14,Jenkins_thermshift,JEN16} as a demonstration of the use of the perturbation theory. We compare the results with results from another class \code{polarizability}, also built on our Software Atom, that computes the susceptibility tensor more directly, basically by applying state-vector perturbation theory.

To showcase the flexibility of the Software Atom, the main program concludes with a small example about a particularly simple steady state that emerges as a result of optical pumping in a constant magnetic field ${\bf B}$. In this optical-pumping example the driving light is not weak, the steady state comes from the direct solution of the equation ${\cal L}\rho_0=0$, and it is unique. We have not done so in our example program, but we could use this steady state as a starting point and investigate perturbatively the response of the system to a weak {\em magnetic\/} field added in a direction perpendicular to the original magnetic field. In this case we would choose a component of the magnetic moment $\mu_\perp$ in some direction perpendicular to ${\bf B}$, use the superoperator corresponding to the operator $-\mu_\perp$ as the superoperator $\cal K$ of the perturbation theory, and the operator $\mu_\perp$ as the operator  $A$ whose expectation value we wish to calculate. Then the relevant member functions of the perturbation class in fact tell us that the assumptions A1-A3 are valid, so in this case $\chi_\perp=\tr(\rho_1\mu_\perp)$  gives us the magnetic susceptibility of the atom in the direction perpendicular to $\bf B$. Likewise, one could study the components of the electric dipole moment induced by the added transverse magnetic field.

\section{Concluding remarks}
We combine a broad framework of quantum mechanics as it is now used in, say, quantum information theory, a quantum-optics view of an atom and its interactions with external fields, and the type facility of C++ to implement these abstractions numerically. The result is small class library that makes it unprecedentedly easy to set up the problem of an atom with angular-momentum degenerate energy levels interacting with a magnetic field and light. We have also presented a perturbative treatment of the state of the atom, and discussed computation of polarizability of an atom as an example. Some of our design decisions may seem counterintuitive, but they are based on long experience in theoretical quantum optics. The unfortunate flip side is that there is a learning curve. We hope we have explained the philosophy behind our C++ classes  in sufficient detail so that it is relatively easy for anyone who runs into multistate issues in quantum-optics style problems to apply them, instead of hand-coding for the specific problem.

 The main point we wish to emphasize here is really a variation of the maxim of the mathematicians, ``if you cannot solve it, generalize it'': By moving to a high enough level of abstraction, it is possible to put together a short piece of C++ code that is both a very flexible and a very general tool for a broad class of physics problems.
 
\section*{Acknowledgements}
Discussions and collaboration on ${}^{87}$Rb with Janne Ruostekoski and Stewart Jenkins are gratefully acknowledged. This work is supported in part by NSF, Grant Number PHY-1401151.

\section*{Appendix: perturbation theory}
In the usual manner we find from Eqs.~\eq{PERTEV} and~\eq{PERTSER} in the zeroth and first order in perturbation theory the equations
\bea
\dot\rho_0 &=& 0\,,\label{ZEROTH}\\
\dot\rho_1 &=&  {\cal L}\rho_1 + e^{\eta t} {\cal K}\rho_0.\label{FIRST}
\eea
Equation~\eq{ZEROTH} holds true since $\rho_0$ was a steady state by assumption A1. Moreover, defining $\rho_1=\tilde\rho_1 e^{\eta t}$, we find for $\tilde\rho_1$ the equation of motion
\beq
\dot{\tilde\rho}_1  =  ({\cal L}-\eta )\tilde\rho_1 +{\cal K}\rho_0\,.
\label{TILDEQM}
\eeq
Obviously ${\tilde\rho}_1$ starts as zero at time $t=-\infty$, and the value we wish to find is $\rho_1(0)=\tilde\rho_1(0)$. The constant $\eta>0$ represents a finite damping, and in effect ensures that there is a solution at $t=0$. The constant $\eta$ is implied in our calculations until otherwise noted, but is not denoted explicitly.

From Eq.~\eq{TILDEQM} we may write an evolution equation over a short time $dt$,
\beq
{\tilde\rho}_1(t+dt) =  \tilde\rho_1(t)+dt[ {\cal L}\tilde\rho_1(t) +  {\cal K}\rho_0]\,.
\eeq
Now, by assumption A2, ${\cal K}\rho_0\in{\cal R}$. Since $\tilde\rho_1$ starts out as the zero operator, the time stepping shows that ${\tilde\rho}_1(t)\in{\cal R}$ at all times. 

Next, introduce the orthogonal projectors $\cal P$ and $\cal Q$ to the subspaces $\cal N$ and ${\cal N}_\perp$. Using ${\cal P}+{\cal Q}=1$ and ${\cal L}{\cal P}=0$, we have from~\eq{TILDEQM}
\bea
\frac{d}{dt}({\cal P}\tilde\rho_1) &=& ({\cal P}{\cal L}{\cal Q})({\cal Q}\tilde\rho_1)+{\cal P}{\cal K}\rho_0\label{PRHOEQ},\\
\frac{d}{dt}({\cal Q}\tilde\rho_1) &=&  ({\cal Q}{\cal L}{\cal Q})({\cal Q}\tilde\rho_1)\label{QRHOEQ} + {\cal Q}{\cal K}\rho_0.
\eea

Equation~\eq{QRHOEQ} leads to the steady-state solution that we again denote by $\rho_1$, namely, $\rho_1\equiv{\cal Q} \tilde\rho_1(0)$. Let us look for a solution to the equation
\beq
{\cal L}\rho_1= - {\cal K}\rho_0;
\eeq
such a solution also automatically satisfies the steady-state version of~\eq{QRHOEQ}.
But since $-{\cal K}\rho_0\in\cal R$ and we require that $\rho_1\in {\cal N}_\perp$, there is a unique solution
\beq
\rho_1 = -\bar{\cal L}{\cal K}\rho_0,
\label{EQU1}
\eeq
where $\bar{\cal L}$ is the pseudoinverse.

In the process there may also arise a correction to the steady state ${\cal P}\tilde\rho_1$, so that the perturbed density operator is, with the parameter $\lambda$ restored,
\beq
\rho(0) = \rho_0 + \lambda[{\cal P}\tilde\rho_1(0) + {\cal Q}\tilde\rho(0)] + {\cal O}(\lambda^2).
\eeq
Now, this density operator is not necessarily normalized to unit trace, so that the expectation value of an operator $A$ satisfying assumption A3 is
\beq
\ave{A} = \frac{\lambda\tr[A{\cal Q}\tilde\rho_1(0)]}{1+\lambda\tr[{\cal P}\tilde\rho_1(0) + {\cal Q}\tilde\rho_1(0)]}+ {\cal O}(\lambda^2) = \lambda\tr(\rho_1A) + {\cal O}(\lambda^2).
\label{EQU2}
\eeq
It may happen that in the final limit $\eta\rightarrow0$ the term ${\cal P}\tilde\rho_1(0)$ diverges, but this is not a problem as the order of the limits, first $\lambda\rightarrow0$ and then $\eta\rightarrow0$, eliminates the divergence.
Equations~\eq{EQU1} and~\eq{EQU2} constitute the final result~\eq{FINALRESULT}. 

\end{document}